\documentclass[12pt,a4paper,dvips]{article}
\usepackage{my_title}

\usepackage{epsfig} \newcommand{\etal}{{\it et~al.}}
\newcommand\GeV{\ifmmode {\mathrm{\ Ge\kern -0.1em V}}\else
\textrm{Ge\kern -0.1em V}\fi}%

\begin{document}

\thispagestyle{empty}

\title{Measurement of the Running of the Electromagnetic Coupling at
LEP}

\author{{Salvatore Mele}\\ \normalsize INFN, Sezione di Napoli, I-80125,
  Napoli, Italy\\ \small {\tt{Salvatore.Mele@cern.ch}}}

\begin{abstract} The study of low-angle and large-angle Bhabha scattering at
LEP gives access to the running of the electromagnetic coupling. Two
recent measurements of the OPAL and L3 collaborations probe the
running of $\alpha$ in the regions $1.8\GeV^2<-Q^2<6.1\GeV^2$ and
$1800 \GeV^{2} < -Q^{2} < 21600 \GeV^{2}$, respectively. The
strategies and the results of these studies are presented. A global
overview is given of the agreement of these and previous L3 findings
with the QED predictions.
\end{abstract}

\vspace{3cm}
\begin{center}
To appear in the Proceedings of the\\ International Europhysics Conference on High Energy
Physics\\
July 21st - 27th 2005,  Lisboa, Portugal
\end{center}

\section{Introduction}

\begin{figure}[bh]
  \begin{center}
    \includegraphics[width=\textwidth]{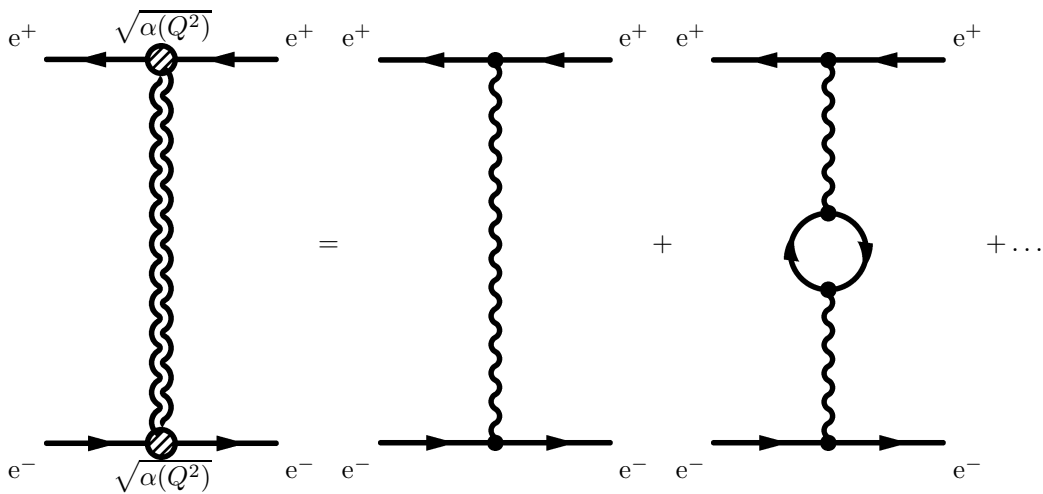}
    \caption{$t$-channel Feynman diagrams contributing to Bhabha
           scattering and the phenomenon of vacuum-polarisation.  The
           sum of all diagrams including zero, one, two or more
           vacuum-polarisation insertions is denoted by the diagram to
           the left with the double-wavy photon propagator, with an
           electromagnetic coupling $\alpha(Q^2)$.}
  \end{center}
  \label{fig:1}
\end{figure}

A remarkable feature of quantum field theory is the dependence of
coupling constants on the energy scale of the processes in which their
corresponding interactions occur. In QED, large momentum-transfers
probe virtual-loop corrections to the photon propagator, as sketched
in Figure~\ref{fig:1}, inducing a dependence of the electromagnetic
coupling, $\alpha$, on the squared momentum transfer, $Q^2$. This
evolution, or {\it{running}}, of $\alpha$ is parametrised
as~\cite{ref:running}:
\begin{equation}
  \alpha(Q^{2}) = \frac{\alpha_0}{1-\Delta\alpha(Q^{2})},
  \label{eq:alfa}
\end{equation}
where $\alpha_0$ is the fine-structure constant. This is measured with
high accuracy in solid-state processes and via the study of the
anomalous magnetic moment of the electron to be $1/\alpha_0 =
137.03599911\pm0.00000046$~\cite{ref:codata}. The contributions to
$\Delta\alpha(Q^{2})$ from lepton loops are precisely predicted, while
those from quark loops are difficult to calculate due to
non-perturbative QCD effects. They are estimated using
dispersion-integral techniques. At the scale of the Z-boson mass, a
recent calculation yields $\alpha^{-1}(m_{\rm Z}^{2}) = 128.936 \pm
0.046$~\cite{ref:burkhardt_new}.

Bhabha scattering at $\rm e^+e^-$ colliders, $\rm e^+e^-\rightarrow
e^+e^-$, constitutes a unique window on the vacuum-polarisation
insertions of Figure~\ref{fig:1}. Its measurement gives access to the
running of $\alpha$ in the {\it space-like} region, $Q^2<0$.  The
four-momentum transfer in Bhabha scattering is measured with high
precision through its dependence on the squared centre-of-mass energy,
$s$, and on the scattering angle, $\theta$: $Q^{2} = t \simeq
-s(1-\cos\theta)/2$.  Low-angle, $\theta\sim 1^\circ-3^\circ$, and
large-angle, $\theta\sim 20^\circ-90^\circ$, Bhabha scattering
correspond to the low-$Q^2$ and high-$Q^2$ regimes, respectively.

A {\it caveat} is in order when discussing the extraction of
information on the running of $\alpha$ from Bhabha scattering. The
cross section of this process can be written as:
\begin{displaymath}
  {{\rm d}\sigma \over {\rm d}t} = {{\rm d}\sigma^{(0)} \over {\rm
  d}t} \left( {\alpha(t) \over \alpha_0}\right )^2
  (1+\varepsilon)(1+\delta_\gamma)+\delta_{\rm Z}, 
\end{displaymath}
\begin{equation}
  {\rm
  with\,\,\, tree-level\,\,\, cross\,\,\, section} {{\rm
  d}\sigma^{(0)} \over {\rm d}t} = { 4 \pi \alpha_0^2 \over t^2 }.
  \label{eq:sigma}
\end{equation}
Naively, one could imagine inserting the measured cross section in the
left-hand side of Equation~\ref{eq:sigma}, and incorporate the knowledge of
the $s$-channel contributions, $\delta_\gamma$ and $\delta_{\rm Z}$,
and of the radiative correction, $\varepsilon$, to extract a value for
$\alpha(t)$. This argument is unfortunately flawed as the measurement
of the cross section requires knowledge of the integrated luminosity.
At LEP this is estimated by measuring events from low-angle Bhabha
scattering and assuming its cross section which, in turns, depends on
$\alpha(t)$. Therefore, the only information which is experimentally
accessible concerns the evolution of $\alpha$ over a $Q^{2}$
range. Two recent measurements of the running of $\alpha$ are
discussed in the following, along with some previous measurements,
and, finally, a combined overview.

\section{Previous measurements}

The L3 collaboration first established the running of $\alpha$ in the
range $2.1\GeV^2<-Q^2<6.2\GeV^2$~\cite{l3-197} by comparing event
counts from low-angle Bhabha scattering in different regions of its
luminosity monitor, with a result:
\begin{equation}
  \alpha^{-1}(-2.1 \GeV^{2}) - \alpha^{-1}(-6.2 \GeV^{2}) = 0.78 \pm
  0.26,
  \label{eq:l3old1}
\end{equation}
where the uncertainty combines statistical and systematic
uncertainties.

The running of $\alpha$ in large-angle Bhabha scattering was first
investigated by the VENUS Collaboration at TRISTAN in the range
$100\GeV^2<-Q^2<2916\GeV^2$~\cite{ref:venus}. Later, the L3
Collaboration studied the same process at $\sqrt{s}=189\GeV$ for
scattering angles $20^\circ-36^\circ$, probing the range
$12.25\GeV^2<-Q^2<3434\GeV^2$~\cite{l3-197}, finding:
\begin{equation}
\alpha^{-1}(-12.25 \GeV^{2}) - \alpha^{-1}(-3434 \GeV^{2}) = 3.80 \pm
1.29,
  \label{eq:l3old2}
\end{equation}
where the uncertainty comprises statistical and systematic sources.

\section{Precision measurement at low \boldmath$Q^2$}

\begin{figure}
  \begin{center}
      \includegraphics[width=\textwidth]{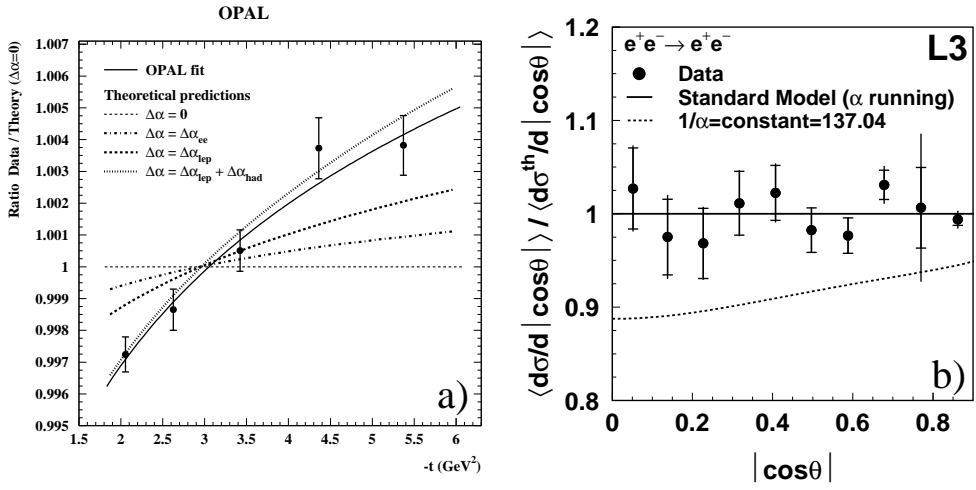}
    \caption{Experimental results proving the running of $\alpha$ at
    a) low and b) high $Q^2$. a) The ratio of event counts in
    different radial regions of the OPAL luminometer, corresponding to
    different values of $t$, for data and a Monte Carlo simulation
    with $\alpha(t)=\alpha_0$, from Reference~\cite{ref:opal}. The
    data favour the QED prediction of $\Delta\alpha(t)$ and exclude
    all other scenarios. Only statistical uncertainties are shown. b)
    The ratio of centre-of-mass-averaged differential cross section of
    for large-angle Bhabha scattering measured by L3 divided by the
    theoretical predictions, from Reference~\cite{ref:l3}. The inner error
    bars denote the statistical uncertainties, the outer the
    combination of statistical and systematic uncertainties. The data
    exclude the scenario $\alpha(t)=\alpha_0$.}
    \label{fig:2}
  \end{center}
\end{figure}

The OPAL collaboration recently performed a high-precision study of
the running of $\alpha$ at low-$Q^2$~\cite{ref:opal} using data
collected with their luminometer. This consisted of tungsten absorber
and 32-pad silicon detectors, covering a polar angle
$1.4^\circ-3.3^\circ$. The analysis relies on 10 million high-energy
back-to-back cluster pairs originating from low-angle Bhabha
scattering at $\sqrt{s}=m_{\rm Z}$, corresponding to a momentum
transfer $1.8\GeV^2<-Q^2<6.1\GeV^2$. The $t$ spectrum of the events is
investigated to extract information on $\alpha(t)$. Data are divided
in five $t$ bins and compared with the BHLUMI Monte
Carlo~\cite{ref:bhlumi}, as shown in Figure~\ref{fig:2}a. The ratio
between data and Monte Carlo counts for the hypothesis
$\alpha(t)=\alpha_0$ is fitted with the function $a+b\ln(t/t_0)$,
where $t_0=-3.3\GeV^2$ is the mean value of $t$ in the data sample and
the parameter $b$ is related to the running of $\alpha$ as
\begin{equation}
\Delta\alpha(t_2)-\Delta\alpha(t_1)\approx {b\over 2} \ln {t_2 \over
t_1}.
\label{eq:b}
\end{equation}
Three hypotheses are considered and found to be completely excluded by
the data: $\alpha(t)=\alpha_0$; a running of $\alpha$ induced only by
electron virtual-loops; a running of $\alpha$ induced only by lepton
virtual-loops. The QED hypothesis of a running with virtual loops of
both leptons and quarks fits the data best. The result of a fit for
$b$, through Equation~\ref{eq:b}, yields:
\begin{equation}
\Delta\alpha(-6.1\GeV^2)-\Delta\alpha(-1.8\GeV^2) = (440 \pm 58 \pm 43
\pm 30) \times 10^{-5},
\label{eq:opal}
\end{equation}
in excellent agreement with the QED prediction of $460 \times 10
^{-5}$. The first uncertainty is statistical, the second systematic
--dominated by the simulation of the detector material and by the
reconstruction of the radial coordinate of the clusters-- and the
third theoretical.

This measurement proves the running of $\alpha$ at low-$Q^2$ with a
significance of $5.6\sigma$, and, for the first time, establishes the
hadronic contribution to the running with a significance of
$3.0\sigma$ as:
\begin{equation}
\Delta\alpha_{had}(-6.07\GeV^2)-\Delta\alpha_{had}(-1.81\GeV^2) = (237
\pm 58 \pm 43 \pm 30) \times 10^{-5}.
\end{equation}

\section{First measurement at large \boldmath$Q^2$}

The L3 collaboration measured the differential cross section for
Bhabha scattering for scattering angles in the range
$26^\circ-90^\circ$. About 40\,000 events are selected as back-to-back
clusters in the high-resolution BGO electromagnetic calorimeter with
matched tracks, at $\sqrt{s}=109-209\GeV$, corresponding to $ 1800
\GeV^{2} < -Q^{2} < 21600 \GeV^{2}$. The 80 measured values of the
cross section, for each of ten angular ranges and eight centre-of-mass
energies, are compared with the predictions of the BHWIDE Monte
Carlo~\cite{ref:BHWIDE} to extract information on the running of
$\alpha$~\cite{ref:l3}.  Figure~\ref{fig:2}b compares the data and the
predictions for the centre-of-mass-averaged cross sections.  The
running of $\alpha$ is parametrised as:
\begin{equation}
  \alpha(Q^{2}) = \frac{\alpha_0}{1-C\Delta\alpha(Q^{2})}.
  \label{eq:runC}
\end{equation}
The hypothesis $\alpha(Q^2)=\alpha_0$, corresponding to $C=0$, is
completely excluded, whereas the data are in excellent agreement with
the running predicted in QED, corresponding to $C=1$. A fit determines
\begin{equation}
C= 1.05 \pm 0.07 \pm 0.14,
\label{eq:c}
\end{equation}
where the first uncertainty is statistical and the second systematic,
dominated by theoretical uncertainties with some contribution from the
detector modelling.

\section{Two textbook plots?}

\begin{figure}[th]
  \begin{center}
    \begin{tabular}{c}
      \includegraphics[width=0.7\textwidth]{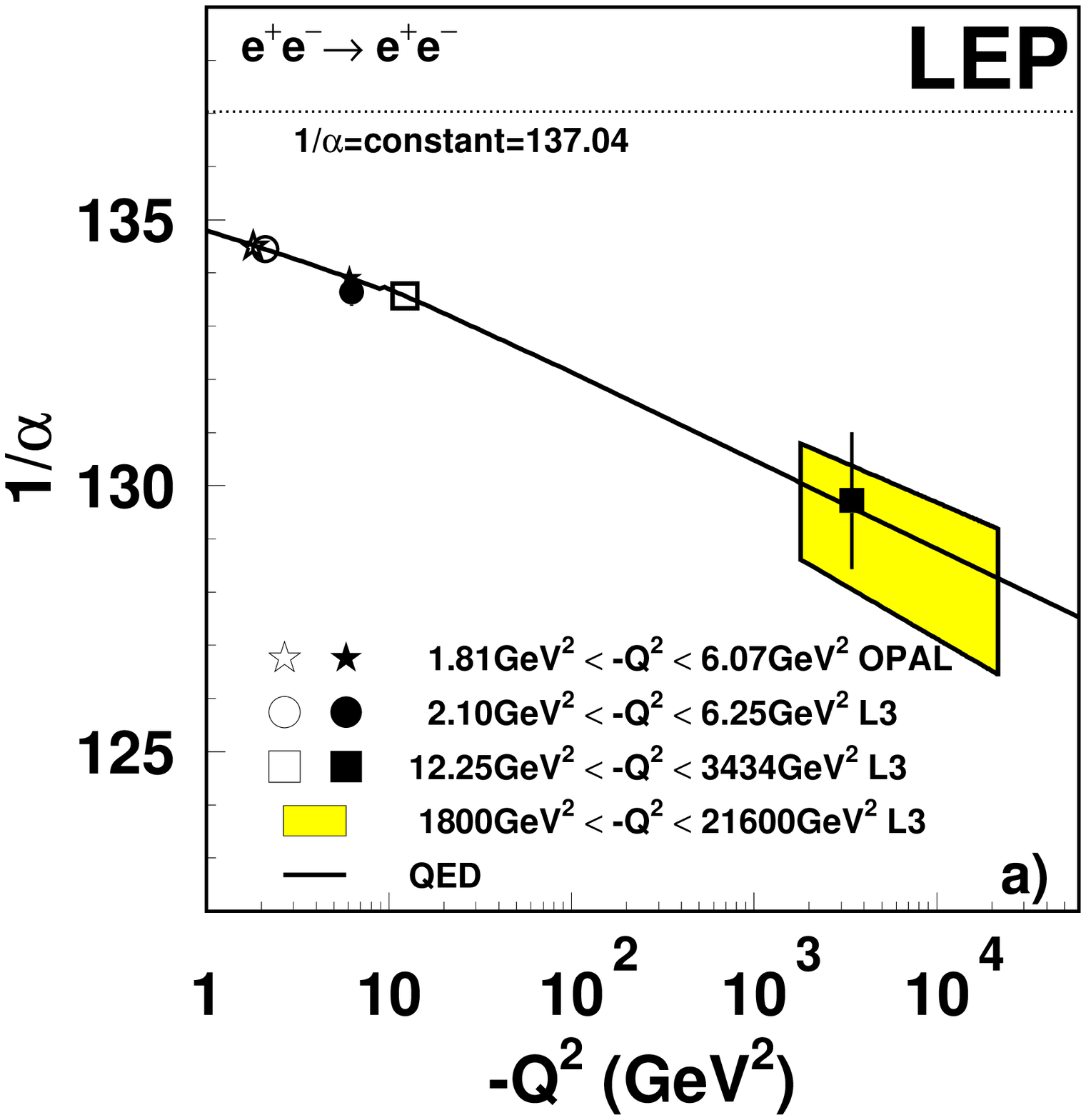}\\
      \includegraphics[width=0.7\textwidth]{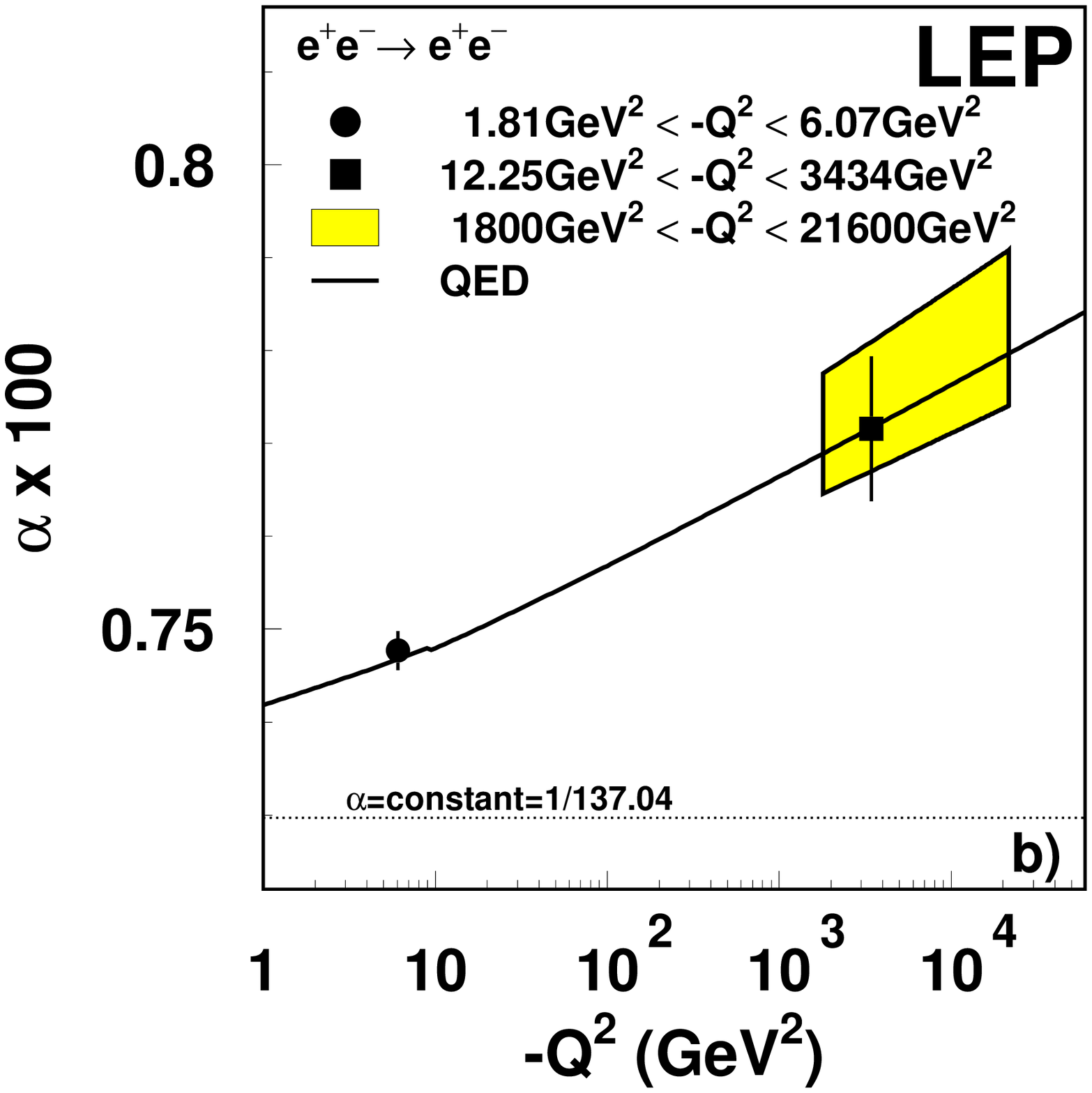}\\
    \end{tabular}
    \caption{Comparison of LEP results on the measurement of the
    running of the electromagnetic coupling with QED predictions. The
    treatment of data and the meaning of the symbols is discussed in
    the last section of the text.}
    \label{fig:3}
  \end{center}
\end{figure}

Figures~\ref{fig:3}a and~\ref{fig:3}b summarise the LEP results on the
running of the electromagnetic coupling. In both figures, the L3
measurement at $ 1800 \GeV^{2} < -Q^{2} < 21600
\GeV^{2}$~\cite{ref:l3} is represented as a yellow band, obtained by
inserting the measured value of $C$ from Equation~\ref{eq:c} into
Equation~\ref{eq:runC} and assuming the QED description of
$\Delta\alpha(Q^2)$ of Reference~\cite{ref:burkhardt_new}.

In Figure~\ref{fig:3}a, the OPAL measurement at
$1.8\GeV^2<-Q^2<6.1\GeV^2$, Equation~\ref{eq:opal}, and the L3
measurements at $2.1\GeV^2<-Q^2<6.2\GeV^2$ and
$12.25\GeV^2<-Q^2<3434\GeV^2$, Equations~\ref{eq:l3old1}
and~\ref{eq:l3old2}, respectively, are represented with a pair of
symbols each. The empty symbol represents the value of
$\alpha^{-1}(Q^2)$ fixed at the lower end of each $Q^2$ range with
Equation~\ref{eq:alfa} and the QED description of $\Delta\alpha(Q^2)$
of Reference~\cite{ref:burkhardt_new}. The full symbol represents the
values of $\alpha^{-1}(Q^2)$ at the higher end of each $Q^2$ range
extracted from these fixed values and the measurements in
Equations~\ref{eq:opal},~\ref{eq:l3old1} and~\ref{eq:l3old2}.

Figure~\ref{fig:3}b goes one step further, anchoring the lower end of
 each $Q^2$ range by using the L3 measurement of $C$ at $ 1800
 \GeV^{2} < -Q^{2} < 21600 \GeV^{2}$ and assuming it also describes
 the running of $\alpha$ for lower values of $Q^2$.  First, the L3
 measurement at $2.1\GeV^2<-Q^2<6.2\GeV^2$ and the OPAL measurement at
 $1.8\GeV^2<-Q^2<6.1\GeV^2$ are combined in a single measurement:
 $\alpha(6.1\GeV^2)-\alpha(1.8\GeV^2)=(363\pm 52)\times 10^{-7}$. The
 value of $\alpha(1.8\GeV^2)$ is then fixed by using the measured
 value of $C$ from Equation~\ref{eq:c}, the evolution expected from
 Equation~\ref{eq:runC} and the QED description of $\Delta\alpha(Q^2)$
 of Reference~\cite{ref:burkhardt_new}. Finally, the value of
 $\alpha(6.1\GeV^2)$ is extracted by using the fixed value of
 $\alpha(1.8\GeV^2)$, with an additional uncertainty which follows
 from the 14\% uncertainty on $C$. A similar procedure is followed to
 extract the value of $\alpha(-3434 \GeV^{2})$ from
 Equation~\ref{eq:l3old2}.

Both Figures present an excellent agreement with the QED predictions
of Reference~\cite{ref:burkhardt_new}, represented by the solid line:
two textbook plots!

\end{document}